\documentclass[12pt]{iopart}
\usepackage{iopams}  
\usepackage[T1]{fontenc}
\usepackage[latin9]{inputenc}
\usepackage{graphicx}
\usepackage{esint}
\usepackage{epsfig}%
\usepackage[usenames,dvipsnames]{color}
\usepackage{mathrsfs}
\usepackage{amssymb}
\usepackage{amsbsy}
\usepackage[abs]{overpic}

\begin{document}

\title{Rate of tunneling nonequilibrium quasiparticles\\ in superconducting qubits}

\author{Mohammad H. Ansari}

\address{Kavli Institute of Nanoscience Delft, \\ Delft University of Technology, 2628 CJ Delft, The Netherlands}
\ead{m.h.ansari@tudelft.nl}
\vspace{10pt}
\begin{indented}
\item[]\today
\end{indented}

\begin{abstract}
In superconducting qubits the lifetime of quantum states cannot be prolonged arbitrarily by  decreasing temperature. At low temperature quasiparticles tunneling between electromagnetic environment and superconducting islands takes the condensate state out of equilibrium due to charge imbalance. { We obtain the tunneling rate from a phenomenological model of non-equilibrium, where nonequilibrium quasiparticle tunnelling stimulates a temperature-dependent chemical potential shift in superconductor.  As a result we obtain a non-monotonic behavior for relaxation rate as function of temperature.} Depending on the fabrication parameters for some qubits the lowest tunneling rate of nonequilibrium quasiparticles can take place only near the onset temperature below which nonequilibrium quasiparticles dominate over equilibrium one. Our theory also indicates that such tunnelings can influence the probability of transitions in qubits through a coupling to the zero-point energy of phase fluctuations.
\end{abstract}

\maketitle

Superconducting integrated devices are among the leading candidates for quantum information processing with meticulous state control \cite{{Insight},{Lucero:2012ex}}. The primary challenge with these devices is the fragility of their quantum states, which makes them vulnerable to various types of decoherence mechanisms.  One of the most important phenomenon to be envisioned as the origin of persistent decoherence at low temperature in superconducting qubits, is the tunneling of nonequilibrium quasiparticles \cite{{Matveev:1993x},{Martinis_2005},{Sendelbach:2008kr}}.  

At low temperature, equilibrium quasiparticle populations should get completely depleted. In practice, however, some nonequilibrium quasiparticles are observed to stimulate qubit energy relaxation time at  $ 10^{2}$ micro-seconds, which is not enhanced by further cooling down \cite{{Houck:2008je},{Bertet:2005cg},{Wang:2009ja},{deVisser:2011cu},{Shaw:2008il},{Bal},{Riste:2012vt},{Martinis:2009uq}, {Paik}, {Corcoles}}. In fact, the coupling between qubit and its surrounding electromagnetic environment stimulates  nonequilibrium quasiparticles leakage between superconducting islands and reservoirs, \cite{{Koch:2007wr},{Lutchyn:2006ei},{Martinis:2009uq},{Martinis:2009tt},{Scherier2008},{Joyez:1994ca},{deVisser:2011cu},{Shaw:2008il},{Bal},{Riste:2012vt},{Paik}, {Corcoles}}.  This causes superconducting qubits to  stochastically carry variable energy levels in a course of time, which makes their complex conductivity variable. Eventually the accuracy of any computations performed longer than the nonequilibrium quasiparticle relaxation time should be degraded in such qubits \cite{{Catelani:2012we},{Catelani:2011cf},{Shaw:2008il}}. One way to obtain the relaxation time is to numerically obtain the nonequilibrium energy distribution in the superconducting island from electron-phonon scattering rates, see \cite{{Martinis:2009uq},{Martinis:2009tt}}, however here we do not take this path.

     This paper is inspired by the charge imbalance effect in a superconducting island upon the tunneling of quasiparticles \cite{{tinkham72},{Tinkhamsolo72},{tinkhambook}}. The theory we study here for interaction between qubit and environment is the so-called $P(E)$ theory extended to strong interactions \cite{Ansari:2012}.  The parity imbalance in  superconducting islands as a result of tunneling from environment takes chemical potential of the condensate out of equilibrium, from which we obtain nonequilibrium quasiparticles rate of tunneling.   We also obtain a relation between nonequilibrium quasiparticle  density and the onset temperature at which deviation from equilibrium tunnelling takes place.  
     
     We finally study how nonequilibrium quasiparticle tunneling affects transitions between qubit states and realize that such effects can be severe.  We provide a general scheme for determining the strength of such intereferences. Only in qubits with the Josephson energy exceeding its capacitive energy the tunneling of quasiparticles may not lead to state transition in qubit.

Upon  tunneling of a quasiparticle in superconducting region, the condensate comes out of equilibrium. To understand this, consider two identical superconducting islands, each in odd parity state. Without loss of generality we can consider a single excess quasiparticle at odd parity states. Now we make the two islands closer to touch each other. The two excess quasiparticles in the combined system after exchang phonons and eventually recombine such that finally there remains no unpaired quasiparticle.  This indicates that an odd parity state in a superconducting region is not a true thermodynamic equilibrium state because if it was so, the combined island would have preferred to maintain the initial states of its subsystem.

Consider a superconducting island whose ground state is formed by pairs of electrons with equal and opposite momentum and spins, i.e. Cooper pairs. { A Cooper pair may occupy the momentum pair state  $(-\mathbf{k},\mathbf{k})$ with probability} $v_k^2=(1-\xi_k/E_k)/2$, or  may not do so (with the probability  $u_k^2=1-v_k^2$).   Now consider a single-particle excitation with electronic charge $e$ and mass $m$ tunnels from outside  into the island; Once arrives at the island it occupies an energy level $E_k= \sqrt{\xi_k^2+ \Delta^2}$ with $\xi_k$  1-particle kinetic energy relative the Fermi energy and $\Delta$ the superconducting energy gap.  
 The quasiparticle can occupy the energy state $E_k$ if either in the lack of Cooper pairs it occupies the momentum state $\mathbf{k}$ (with the Fermi probability $f_{\mathbf{k}}$), or if in the presence of Cooper pair it unoccupies the momentum state $\mathbf{-k}$. The induced charge is  $q/e= \sum_k u_k^2 f_{\mathbf{k}}+v_k^2 (1-f_{-\mathbf{k}})$. This charge can be seen as the sum of superfluid charge  $q_s/e\equiv \sum_k v_k^2$ and quasiparticle charge $q_{\textup{qp}}/e\equiv  \sum_{\mathbf{k}} u_k^2 f_{\mathbf{k}} - v_k^2 f_{-\mathbf{k}}$.   
 
 In thermal equilibrium the occupation probability is the Fermi function $f^0_{\mathbf{k}}$, which is an even function in momentum reflection$f^0_{\mathbf{k}}=f^0_{-\mathbf{k}}$.   Therefore, the charge of  tunneling equilibrium quasiparticles turns out to be  $q_{\textup{qp}}= \sum_k q_k f^{0}_{\mathbf{k}}$, defining  $q_k\equiv e (u_k^2-v_k^2) = e \xi_k/E_k$.  Therefore, the total charge imbalance in the island will be  maximally electron-like (with charge $e$) for fast electrons $|\mathbf{k}| \gg |\mathbf{k}_F|$ (i.e. $\xi_{k} \gg \Delta$) or  minimally hole-like (with charge $-e$) for slow particle of $|\mathbf{k}| \ll |\mathbf{k}_F|$.  Summing over all such quasiparticles the total charge imbalance in thermal equilibrium completely vanishes.  Consequently, equilibrium quasiparticles are depleted as superconductors are electrically neutral.  

In nonequilibrium case, the occupation probability is  $f_{\mathbf{k}}$, which with respect to momentum reflection from Fermi surface may contain both even and odd terms.   The even mode (i.e. $f_{\mathbf{k}}^0$) produces equally both electron-like and hole-like quasiparticles which may contribute only in shifting the superconducting energy gap. This mode is labelled as ``longitudinal'' or ``temperature mode'' of quasiparticles.  However, odd-momentum terms inject more charge into one side of electron-like or hole-like branches and eventually stimulate a charge imbalance in the region, $\sum q_k (f_{\mathbf{k}}-f_{\mathbf{k}}^0)$.  This charge imbalance can be directly measured for instance in a voltage biased Josephson junctions, where the induced current is made of the Giaever current and  a (bias-independent) part that carries only the charge imbalance. The latter leads to a measurable steady-state voltage, (see \cite{{tinkham72},{Tinkhamsolo72}}). The excess current provides valuable information about the tunnelling of nonequilibrium quasiparticles, such as their non-local nature \cite{{Hubler:2010fw},{PhysRevLett.109.087004}},  however we do not discuss such effects here as they minimally contribute to our results. 

In general there is no physical reason to prevent us from considering that a number of quasiparticles may tunnel in (or out) of a superconducting island. Let us consider  one pair of them at the island. They may recombine by exchanging  phonons, however in mesoscopic superconducting qubits made of standard superconductors such recombination process takes place over a few milliseconds, which is almost 2 orders of magnitude longer  than the time needed for the quasiparticles to reach a thermal equilibrium with the island \cite{{Parker:1975ji},{Giazotto}}.  This poses a bottleneck over the recombination process and allows the accumulation of quasiparticles  in the island before they recombine.  However, in practice such abundance of charge in superconducting region does not influence the macroscopic quantum dynamics of a Josephson junction. This is due to charge neutrality of the superconducting region and ensures that the abundance of quasiparticle charge is compensated by superfluid charge  such that the new superfluid charge should be shifted to  $q_s-\delta q_s$.  Looking at above definitions, one can expect the suppression of superfluid charge can be due to reduction in the probability of occupation by Cooper pairs (or equivalently from the rise of kinetic energy.)  This indicates that in superconducting region the chemical potential is shifted from $\mu$ to $\mu-\delta \mu$.  In brief, the  abundance of nonequilibrium unpaired quasiparticles effectively behaves as a Fermi gas with shifted occupation probability  $f_0(\epsilon, \mu-\delta \mu)$ \cite{{Owen:1972gk},{Aronov},{Elesin}}.   The use of this function has been previously used in charge qubit and Cooper pair box \cite{{Shaw:2008il},{Palmer:2007kv},{lutchyn2005}}, however here we use this concept for all qubits.

The density of nonequilibrium quasiparticles number in a superconducting region is  $n_{nqp}= \int d\epsilon \rho(\epsilon) [ f-f_{0}]   $, which using the chemical potential shift can be simplified to:

\begin{equation}
\label{eq. n_nqp}
n_{nqp}= \int d\epsilon \rho(\epsilon) \left[ f_0(\epsilon, \mu- \delta \mu)-f_{0}(\epsilon, \mu) \right],
\end{equation}
where the density of states is $\rho(\epsilon)=\rho_0\Theta(|\epsilon|-\Delta)\ |\epsilon|/\sqrt{\epsilon^2-\Delta^2}$ with the Fermi energy $E_f$, density of states at the Fermi energy $\rho_0$, step function  $\Theta(x)$.  The shifted chemical potential is (for derivation see Appendix A): 

\begin{equation}
\label{eq. chemical pot shift}
\delta\mu=k_B T\ln\left(1+\frac{n_{nqp}\left(1+3k_B T/8\Delta\right)}{\rho_0\sqrt{2\pi\Delta k_B T}}\ e^{\frac{\Delta}{k_B T}} \ \right)
\end{equation}
where $k_B$ is the Boltzman coefficient. 

Note that the  ratio  $ n_{nqp}/  \rho_0 \Delta $ is in fact  the excess quasiparticle number per Cooper pair.   With the increase of the ratio the nonequilibrium effects start to take dominance over equilibrium in the superconducting region.  

The onset temperature for nonequilibrium effects can be evaluated by considering the chemical potential shift of the order of thermal fluctuations $\sim k_BT$.  This helps to provide the nonequilibrium onset temperature: 

\begin{equation}
\label{eq. onset temp}
T_{o}=2 \Delta /(\ln z - \ln (\ln z))k_B
\end{equation}
where $z = 4\pi ( \rho_0 \Delta/n_{nqp})^2$.  

Let us check this relation with a typical qubit. In  Ref. \cite{Martinis:2009uq} the temperature dependence of  quasiparticle tunneling rates for a qubit was reported. From data one can obtain the onset temperature for nonequilibrium effects is at $\sim 0.15$.   Given the parameters related to the qubit made of aluminum with gap $\Delta/k_B=2.1$K,  the Cooper pair density $2.8\times 10^6/\mu\textup{m}^3$, and quasiparticle density of about $10/\mu\textup{m}^3$, our theory predicts the onset temperature is at $T_o=0.16$K, which is almost what has been measured in the experiment.

As discussed when nonequilibrium quasiparticle tunnels through a superconducting island, it induces a chemical potential shift in the region and therefore energetically couples to the qubit performed by the  corresponding islands.  For a charge qubit this is schematically illustrated in Figure (\ref{fig. 1}a) where the first two energy eigenvalues of qubit are plotted with respect to charge gate $n_g$. Depending on the value of $n_g$ different transitions between even and odd parity states may occur.  At $n_g=1/2$ the transition from odd to even parity can take place in relaxation from $C$ to $A$ with qubit frequency as well as from $B$ to $A$ at much lower frequency. However the opposite transition from even to odd parity takes place dominantly from $A$ to $B$ at low frequency {because excitation with higher frequency from $A$ all the way up to  $C$ takes place with much lower rate. These transitions are illustrated in the energy profile of Figure (\ref{fig. 1}b).  

The tunneling of nonequilibrium quasiparticles between superconducting reservoirs and islands, as discussed, stimulates energy exchange between qubit and the quasiparticles, which yields a suppression in the island superfluid charge.  In a course of time this can be seen as fluctuations in the chemical potential of the condensate state, or equivalently  an effective  voltage noise $\delta V(t)$ across the junction. Such a voltage noise induces a phase fluctuations across the junction $\delta \phi(t)/2 = (e/\hbar) \int_{0}^t dt' \delta V(t')$ where  $\phi(t)=\varphi +\delta \phi(t)$ and   $\varphi$ is the semi-classical Josephson phase. 

Using P(E) theory, the  power spectrum of phase fluctuations in the qubit $S(t)=\langle \delta \phi(t) \delta \phi(0) \rangle $ can be obtained from the fluctuation-dissipation theorem \cite{Ingold:880853}.   
To see this we followed in \cite{Ansari:2012}  the RCSJ models and its microscopic
analogue \cite{tinkhambook,Ambegaokar82,Caldeira83,Caldeira81},  where the quasiparticle channel is put in parallel to supercurrent,
Josephson channel, the junction capacitance $C$, and the external impedance
$Z(\omega)$.  In qubits with high quality factor, the effective impedance of junction is restricted to the absorption and emission of Plasma frequency $\omega_{p}=1/\sqrt{CL_{J}}$, i.e. $Z_{\textup{eff}}=(\pi/2C)(\delta(\omega - \omega_p)+\delta(\omega+\omega_p))$, where the  Josephson inductance is $L_{J}=\Phi_{0}/2\pi I_{o}$, critical current $I_{o} $, and the quantum of flux $\Phi_0= 2.07\times 10^{-15} \textup{Wb}$. The noise from resistors gets shunted in the junction at low frequencies; therefore  the electromagnetic environmental effects can be encoded in the transmission of quasiparticle channels.     In Ref. \cite{Ansari:2012} we introduce a parameter called the dimensionless Cooper pair wave impedance $\rho_c=4\pi Z_0/R_K$ with $Z_0=\sqrt{L_J/ C}$ and  resistance quantum $R_K = \hbar/e^2 = 25.8 \textup{k}\Omega$.  From the above definitions one can find that  Cooper pair   wave impedance is $\rho_c=\sqrt{E_c/2 E_J}$. According to Ref. \cite{Ingold:880853},  the vacuum fluctuations induced by the quasiparticle channel in the condensate state is controlled by the wave impedance $S(0)=\rho_c$  across a junction.

The Josephson phase dynamics follows from the low-energy effective Hamiltonian of a qubit interacting with quasiparticles.   In the Appendix B  we introduced the Hamiltonian and determined the transition rate between parity states using the Fermi's golden rule.  Now substituting the nonequilibrium energy distribution  we discussed above,  the explicit  dependence of tunneling rate on temperature, frequency, and phase fluctuations becomes evident (see Eq. (B.6)).     In the limit of a phase qubit with large capacitance junction and the superconducting phase different across the junction $\varphi\approx 0$ the decay rate of nonequilibrium quasiparticles at low temperature turns out to be:
\begin{equation}
\label{eq. rate egeneral}
\Gamma(T) =   \left( \frac{\rho_{c} \Delta}{4E_q } K_{0}\left(\xi\right) + \left(  1+\frac{\rho_{c}}{8}\right)K_{1}\left(\xi\right) \right) \Gamma_o \sqrt{\xi}   \exp(\xi -\frac{\rho_{c}}{2}) ,
 \end{equation}
where $\xi(T)\equiv E_q /2k_BT$ with  qubit energy $E_q$ and $\Gamma_o\equiv  \frac{n_{nqp}}{2 e^2 R_N \rho_0 }\sqrt{\frac{ E_q E_c}{\pi \Delta E_J  }} $, and $K_n(x)$  the Modified Bessel function of the second kind.   

Figure (\ref{fig. 1}c) displays the temperature dependence of the quasiparticle relaxation time (i.e. $1/\Gamma$) in a typical qubit with frequency 1GHz and the ratio $E_J/E_c$ evenly spaced from $5$ to $75$ from bottom to top, respectively.   Dashed line indicate the equilibrium relaxation time for  $E_J/E_c=75$. By {cooling the qubit, its relaxation time makes a rapid decrease in slope from the steep slope increase of equilibrium into the almost stable  regime of nonequlibrium.}  As the nonequilibrium takes effect at low temperature, nonequilibrium  quasiparticles relax much faster (by several orders of magnitude) than what equilibrium quasiparticles do in similar situation.  The inset plot displays the corresponding transition rates of two qubits: the upper one for $E_{J}/E_{c} \sim 1$ and the bottom one for $\gg 1$ and their equilibrium rates in dashed lines. 

In  qubits with $E_J \sim  E_c$ relaxation rate suppresses linearly at low temperature, however as the Josephson energy increases  the tunneling rate becomes slightly enhanced,  e.g. see upper inset of Fig. (\ref{fig. 1}c) .  The  reason is, as we showed in Ref. \cite{Ansari:2012}, the tunneling rate of processes which create charge superpositions (such as charge imbalance phenomenon discussed in here) are suppressed exponentially by zero-point phase fluctuations. Therefore  increasing the Josephson energy makes phase fluctuations decrease and the tunneling rate gets enhanced.  For qubits with $E_J>E_c$, about the onset temperature $T_o$,  nonequilibrium quasiparticles reach to a minimum tunneling rate. The existence of such a minimum has been reported in  a phase qubit in Ref. \cite{{Martinis:2009uq}}.   The  regime of small capacitive energy has also been practically experimented in a transmon qubit where similar monotonic suppression of tunneling rate was observed \cite{{Bal},{Riste:2012vt},{catelani2014}}.   

From eq. (\ref{eq. rate egeneral}) one can find that a minimum in tunneling rate can occur in a qubit with $\rho>\rho_c^*  (T)$ with the critical coupling is $\rho_c^*(0) \approx 3  E_q /\Delta$; (for detail analysis see Appendix C).

\begin{figure}[tb]
\center
\includegraphics[scale=0.25]{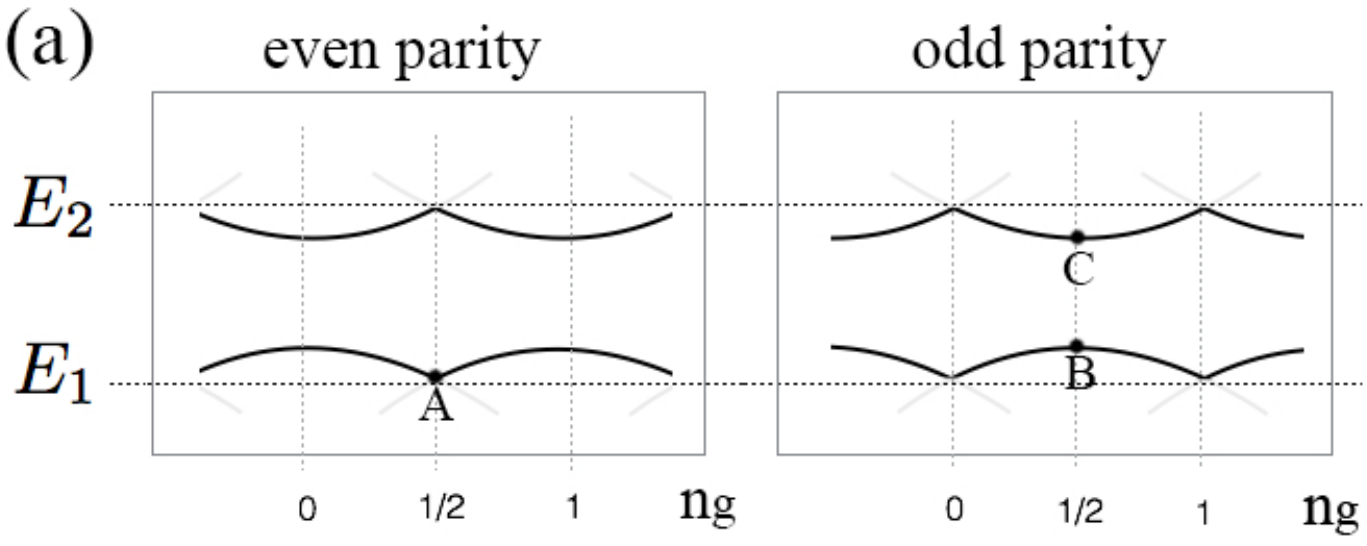}
\includegraphics[scale=0.12]{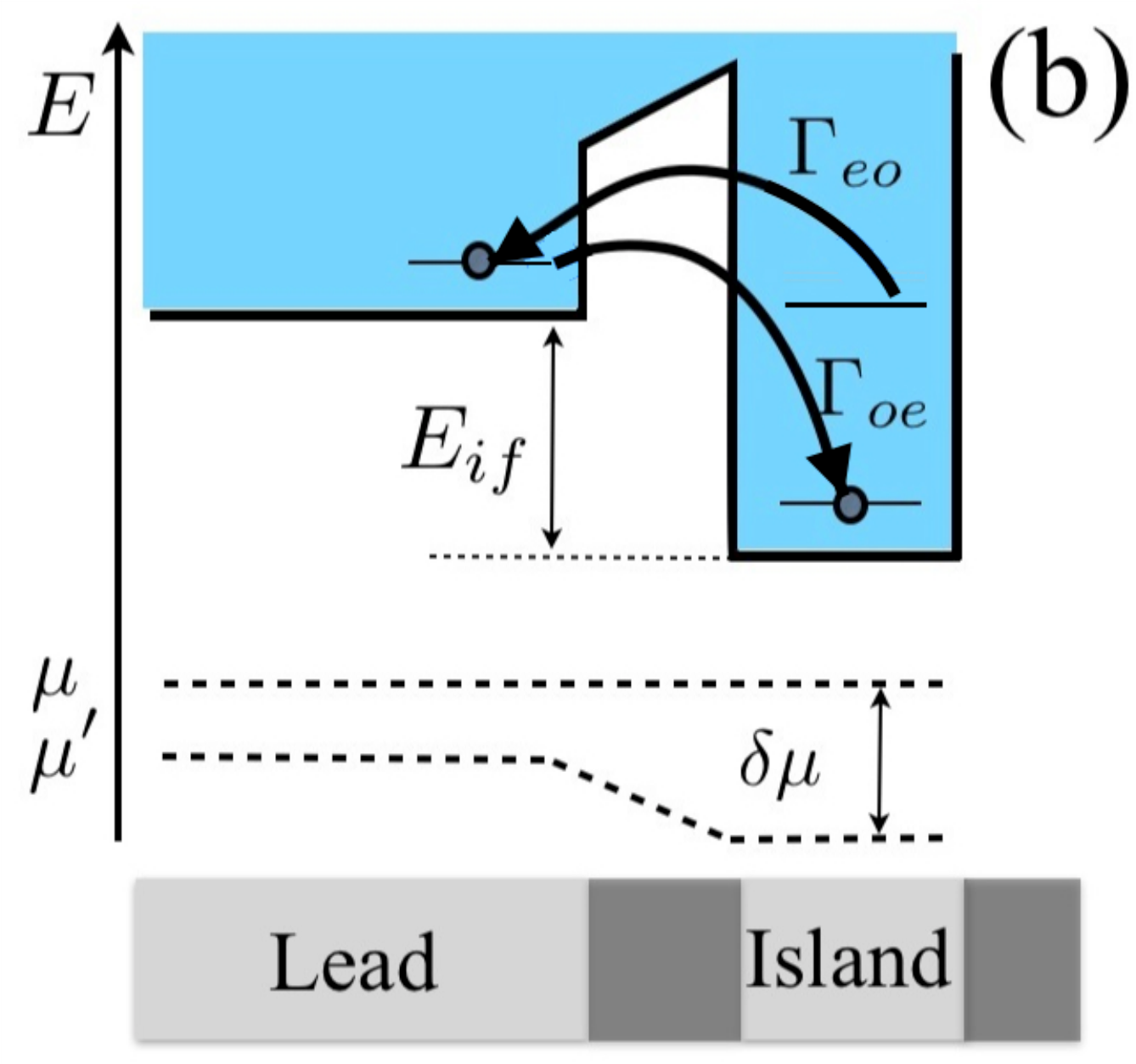}
\begin{overpic}
[scale=1.3,unit=0.5mm]%
{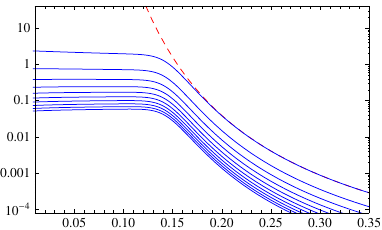}
\put(-5,96){$1/\Gamma$}
\put(-5,88){\footnotesize{(ms)}}
\put(147,-5){$T$(K)}
\put(20,88){(c)}
\put(93,51){\includegraphics[scale=.65]%
{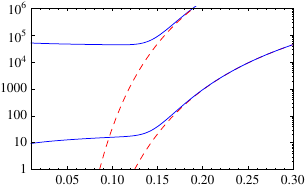}}
\put(90,93){\tiny{$\Gamma$(Hz)}}
\put(150,47){\tiny{T(K)}}
\put(16,8){\includegraphics[scale=.12]%
{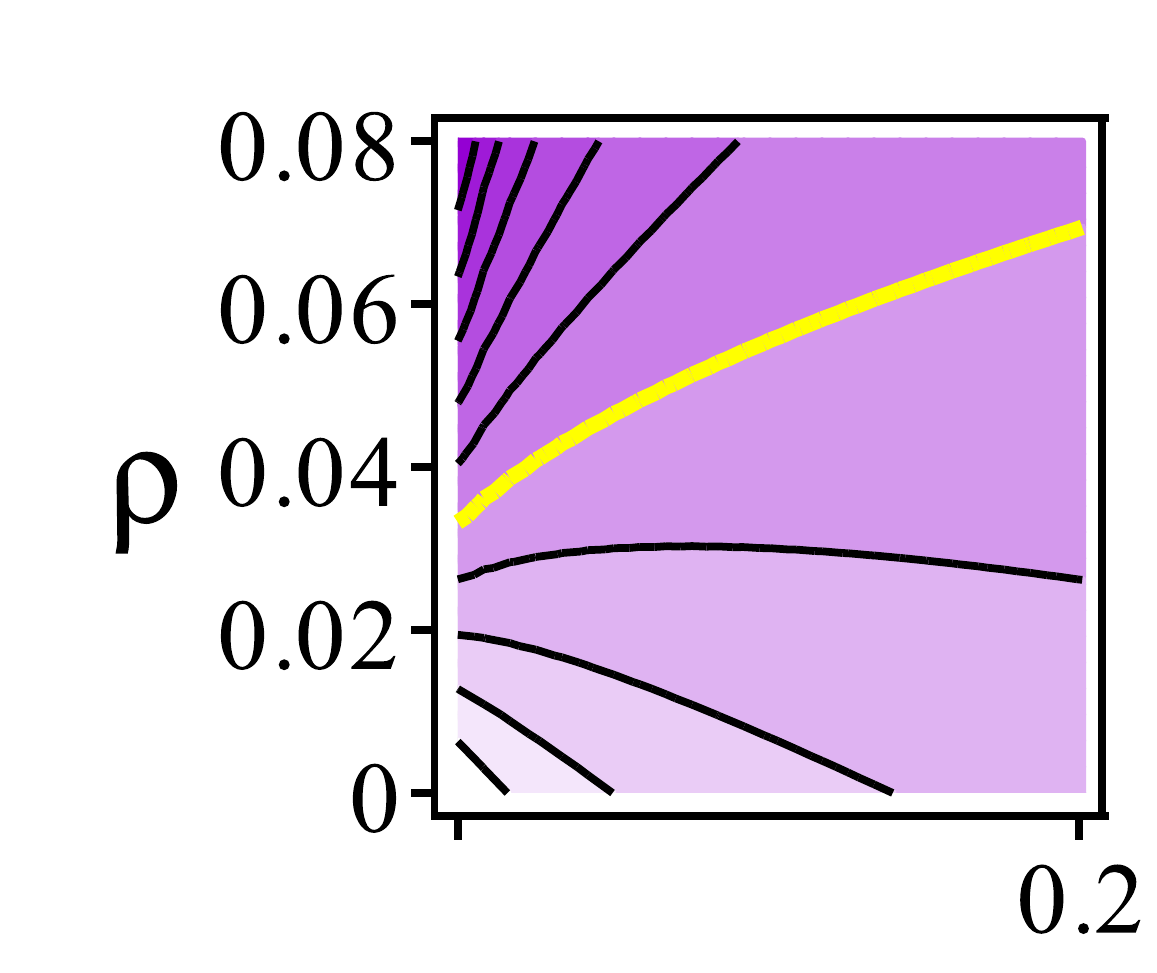}}
\put(61,14){\includegraphics[scale=.051]%
{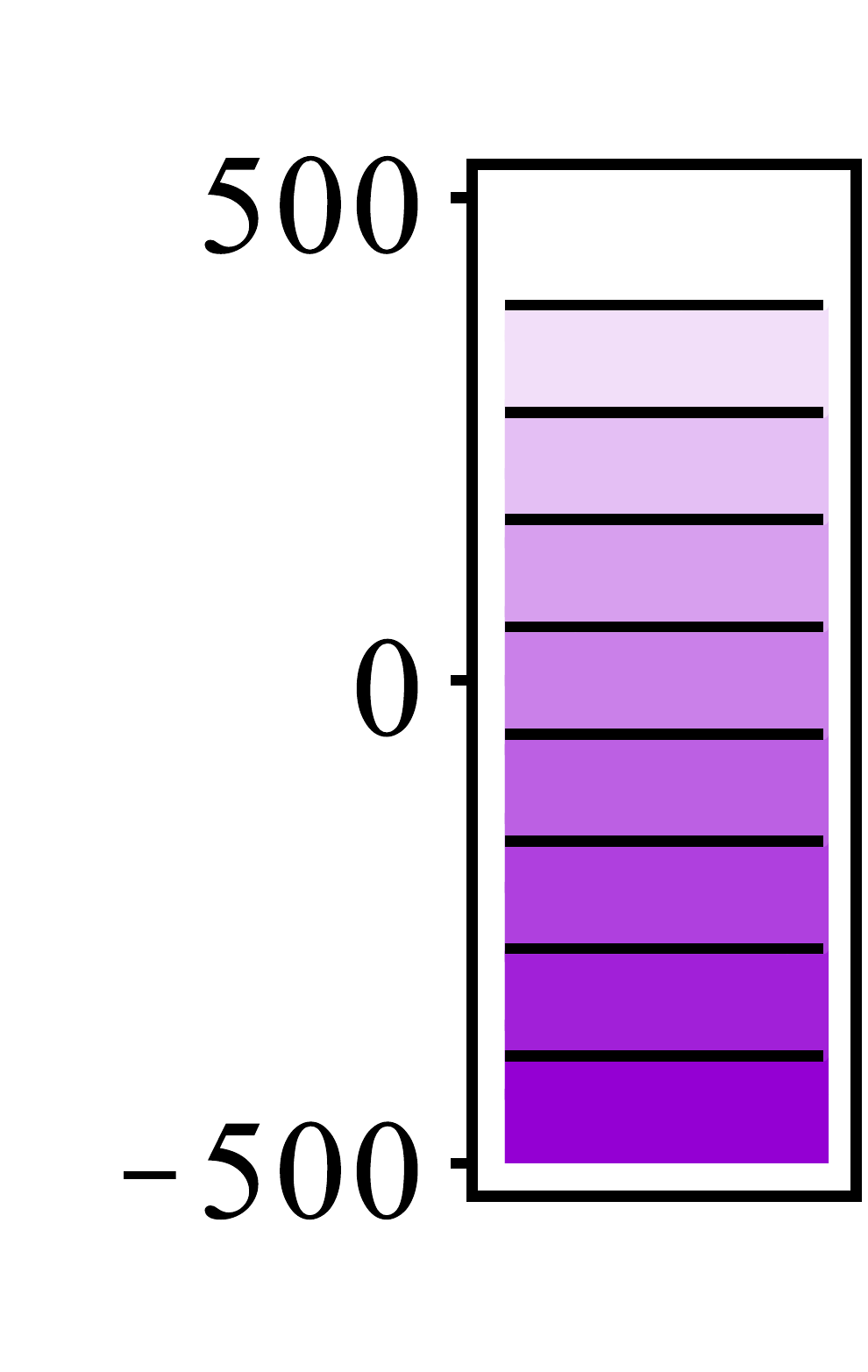}}
\put(63,11){\tiny{T}}
\put(78,24){\tiny{$\frac{d\Gamma}{dT}$}}
\put(51,35){\tiny{$\rho_c^*$}}
\end{overpic}
\caption{(a)  The transition between even and odd parity states; both initial and final states are tuned at a fixed charge gate $n$.  (b) The energy profile at a junction between a large reservoir (lead) and a small island.  The even to odd transition rate $\Gamma_{eo}$ and the opposite one are depicted by arrows from the corresponding initial to final states.  Large transition energy makes the even to odd transitions much more frequent than that in the opposite direction. (c) Temperature dependence of NQP relaxation time for the superconducting energy gap $\Delta=2.1K$,  $R=10\textup{K}\Omega$,  the nonequilibrium quasiparticle density $0.1/\mu m^3$, and the density of states $10^6/ \mu m^3K$.  From top to bottom the junction phase fluctuations is taken from $\rho_{c}=0.08$ to $0.8$ with equal steps.  Upper Insets:  temperature dependence of  relaxation rate in nonequilibrium (solid) and equilibrium (dashed), {where the upper curve is} for $\rho_{c}=0.04$ and the lower one for 1.6; Lower Inset: the slope of NQP relaxation rate as a function of temperature and $\rho$. Minima in the transition rate appears for all $\rho_c> \rho_c^*$ at finite temperature.}
\label{fig. 1}
\end{figure}
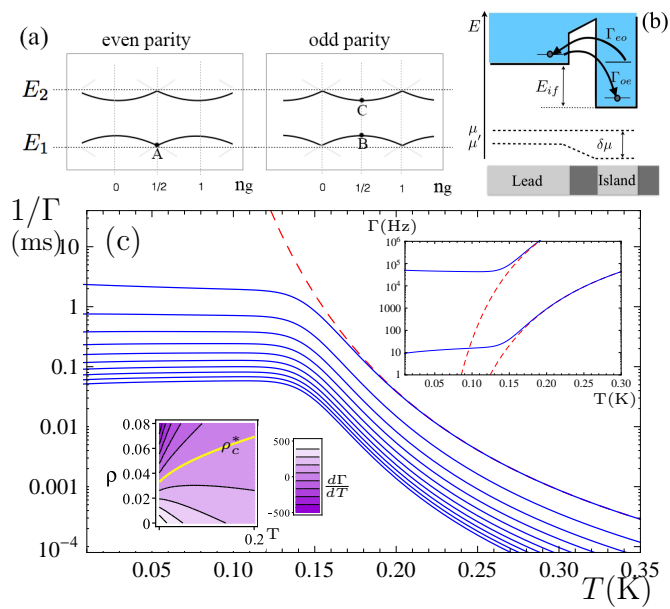

The slow variation of the tunneling rate by temperature makes one to  expect that in contrast to what equilibrium models predict,  the rate of tunneling in nonequilibrium quasiparticles should approach a finite value at zero temperature ($T\sim 0$). In  arbitrary superconducting phase across a junction the complete analysis (see Eq. (B.8) in Appendix) shows the zero temperature limit of the tunneling rate in a junction with arbitrary macroscopic Cooper pair phase $\varphi$:
\begin{equation}
 \Gamma=\frac{\alpha n_{nqp} \exp\left({-\sqrt{\frac{E_c}{8E_J}}}\right) }{2e^{2}\rho_0R_{N}} \ ( 1 -\eta \cos {\varphi} ),
\label{eq. appA final 2}
\end{equation}
defining  $\alpha\equiv \sqrt{\frac{ E_c \Delta}{ 16 E_q  E_J }} \left(1+\sqrt{\frac{E_c}{32E_J}} \right)$ and $\eta\approx 1-\sqrt{E_c/8E_J}$. In the equilibrium limit (i.e. by setting $\delta \mu=0$)  this rate becomes equivalent to the equilibrium rate that is proportional to $ \sqrt{T} \exp \left( -\Delta / k_{B}T \right)$ .

There are important novel results to be understood from eq. (\ref{eq. appA final 2}). It is important to notice that the coefficient  $\eta$ of the $\cos$ term. This term represents the energy transition matrix in a qubit coupled to environment.  { Some models for tunneling at  low quasiparticle energy, such as \cite{{catelani_2011_QpRelax},{Catelani:2012we}}, showed that  $C \to A$ qubit transitions [see Fig. 1a] takes place with a phase dependence rate proportional to $(1- \cos \hat{\varphi})$. In \cite{Catelani:2011cf} a correction to this relation has been discussed - see Eq. 16 in that reference. Recent studies in \cite{{Leppakangas2011},{Leppakangas2012}} also indicated the necessity of a correction to the  qubit transition matrix.  Our model in Eq. (\ref{eq. appA final 2}) determines the phase dependence for $B \to A$ transitions to be $1 -\eta \cos \hat{\varphi} $.   This  shows that not only the qubit transition is phase dependent, but also its interaction with nonequilibrium quasiparticles (through the new parameter  $\eta$) determines how to engineer a superconducting quantum device more protected from such tunnelling. }

Discussion: In this paper  we studied the tunneling of nonequilibrium quasiparticles in superconducting qubit using a model for determining how such tunneling may cause to shift  chemical potential of the condensate.  We determined the onset temperature of nonequilibrium effects. This temperature can serve as a probe quantity to determine the density of nonequilibrium quasiparticles in the device. {Our theory reveals a striking modification to qubit transition affected by low energy quasiparticles, where the phase-dependence of transition matrix shows non-trivial coupling to zero-point phase fluctuations.  This can make a dramatic enhancement or reduction in the qubit interference. This effect is expected to be more visible in large junctions, which are by now most commonly used in macroscopic qubit transitions.} We showed that the onset temperature of nonequilibrium regime $T_o$ is a measurable quantity to determine the density of nonequilibrium quasiparticles in a device.    Moreover, the transition rate in qubit gets affected from the phase fluctuations stimulated by the tunneling in a way that its phase dependence is modified to $1-\eta \cos \phi$.  This modification can provide a basis for studying new effects in superconducting qubits and also to improve the efficiency of quantum control.  There are some fabrication related hints that can be concluded from our theory. In general, as  shown in Fig. (1.c),  the tunneling rate of nonequilibrium quasiparticles is globally suppressed at all temperatures in qubits with larger Josephson energy compared to capacitive energy. Yet, in devices with smaller Josephson energies, although we expect a higher rate of quasiparticles, but the lower rate is reached at the onset temperature $T_0$. Depending on the parameters, sometimes the rate at $T_0$ is the lowest one such that decreasing temperature below it will increase the rate again. This is in fact a pure effect of nonequilibrium. Moreover, the non-trivial phase dependence of transition matrix in qubit, in the form we derived, can result in the exotic situation of getting some energy levels in qubit partially forbidden as a result of phase fluctuations resulted from nonequilibrium quasiparticle tunneling.

Beyond quantum computing, our treatment highlights nonequilibrium effects in small islands in the presence of spurious interactions with environment. 
Among many possibilities for future development of this they we propose that one can determine a method for measuring the rate in a qubit that provides a feedback information on the distribution of nonequilibrium quasiparticle.


\emph{Acknowledgements: }  The author would like to thank  Frank K. Wilhelm-Mauch and  Emily Prichett for helpful discussions and  comments.

\appendix

\section{Chemical potential shift $\delta \mu$}
Using the definition of the nonequilibrium quasiparticle density $n_{nqp}= N_{nqp}/V= \int d\epsilon D(\epsilon) [ f_0(\epsilon, \mu- \delta \mu)-f_{0}(\epsilon, \mu) ]   $  one can use the superconducting density of states in superconductor based on the normal metal density of states $D_{n}$, and in the limit of low temperature finds
\begin{eqnarray}\nonumber
n_{nqp}& =&2\int\left[\frac{E}{\sqrt{E^{2}-\Delta^{2}}}D_{n}\right]\frac{1-X}{1+X+e^{-E/kT}+Xe^{E/kT}}dE\\ \nonumber
&\sim&\sqrt{2\Delta}\frac{(1-X)}{X}e^{-\Delta/kT}D_{n}(E_{f})\int_{\Delta}^{\infty}\left(\frac{1}{\sqrt{E-\Delta}}+\frac{3}{4\Delta}\sqrt{E-\Delta}\right)\frac{1}{e^{\left(E-\Delta\right)/kT}}dE\\ \nonumber
&\sim&\sqrt{2\pi\Delta kT}\left(1+\frac{3kT}{8\Delta}\right)\frac{(1-X)}{X}e^{-\Delta/kT}D_{n}(E_{f})
\end{eqnarray}
where we used the definition $X=\exp(\delta \mu/kT)$. Solving the final equation results into the chemical potential shift.

\section{Rate}

We formulate the mathematical description of quasiparticle tunneling
in an environment. We start from the Hamiltonian \cite{{Martinis:2009uq},{Martinis:2009tt}, {Scherier2008},{Joyez:1994ca},{Catelani:2012we},{Catelani:2011cf},{catelani_2011_QpRelax}}
\begin{equation}
\hat{H}=\hat{H}_{{\rm BCS,1}}+\hat{H}_{{\rm BCS,2}}+\hat{H}_{T}+\hat{H}_{{\rm env}}+\hat{H}_{{\rm env-c}},
\end{equation}
for BCS Hamiltonians in mean field form describing the two electrodes and the tunneling Hamiltonian 
\begin{equation}
\hat{H}_{T}=\sum_{kl}\left(T_{kl}\hat{c}_{k,\sigma,1}^{\dagger}\hat{c}_{l,\sigma,2}+T_{kl}^{\ast}\hat{c}_{k,\sigma,2}^{\dagger}\hat{c}_{l,\sigma,1}\right)\label{eq:tunneling}
\end{equation}

By diagonalize the BCS Hamiltonians through the Bogoliubov transformations, one can rewrite the tunneling Hamiltonian, eq. (\ref{eq:tunneling}) in the following form

\begin{eqnarray}
\hat{H}_{T} & = & \sum_{k,\sigma=\uparrow,\downarrow}\left(T_{kl}|u_{k1}u_{l2}|e^{i\phi/2}-T_{-l-k}^{*}|v_{k1}v_{l2}|e^{-i\phi/2}\right)\nonumber \\
 &  & \times\left(\hat{\gamma}_{k\sigma1}^{\dagger}\hat{\gamma}_{k\sigma2}+\hat{\gamma}_{-k\sigma2}^{\dagger}\hat{\gamma}_{-l\sigma1}\right)+H_{T2}\label{eq. HT}
\end{eqnarray}
where we introduced the phase difference $\phi=\phi_{1}-\phi_{2}$.
The term $\hat{H}_{T2}$ contains operators that change the number
of quasiparticles by two, i.e., contain terms of the structure $\hat{\gamma}\hat{\gamma}$
and $\hat{\gamma}^{\dagger}\hat{\gamma}^{\dagger}$ hence changing
the total number of quasiparticles in the setup, which do not contribute
to the quasiparticle rate --these terms contribute to Josephson and
Andreev processes. 

We now want to evaluate the Fermi's golden rule
rate for a transition that transfers a quasiparticle from  electrode
1 to 2. The relevant matrix element is 
\begin{eqnarray*}
\langle\left(N_{1}-1\right)_{k\sigma},\left(N_{2}+1\right)_{l\sigma}\left|\hat{H}_{T}\right|N_{1k\sigma},N_{2l\sigma}\rangle=\\
T_{kl}e^{i\phi/2}\left|u_{k1}u_{l2}\right|-T_{-k-l}^{*}e^{-i\phi/2}\left|v_{k1}v_{l2}\right|
\end{eqnarray*}

In order to capture the influence of the environment, we apply the
ideas of $P(E)$-Theory \cite{Ingold:880853}. There, the environmental
Hamiltonian is described by an oscillator bath $\hat{H}_{{\rm bath}}=\sum_{n}\omega_{n}\hat{a}_{n}^{\dagger}\hat{a_{n}}$, 
which couples the oscillators linearly to our quasiparticle system $\hat{H}_{env-c}=\hat{N}_{1}e\delta\hat{V}$,
where we assume that only the first reservoir fluctuates - this corresponds
to a specific choice of gauge \cite{Ansari:2012}. The total number and voltage operators
are 
\begin{equation}
\hat{N}_{1}=\sum_{k,\sigma}\hat{c}_{k\sigma1}^{\dagger}\hat{c}_{k\sigma1},\quad\delta\hat{V}=\sum\lambda_{i}\left(\hat{a}_{i}+\hat{a}_{i}^{\dagger}\right).
\end{equation}

In the limit of single boson exchange in a large capacitance junction,  \cite{Ansari:2012}, one can rewrite the quasiparticle contribution as
\begin{eqnarray}
\nonumber 
\vec{\Gamma}_{1} & = & \frac{1}{2 e^{2}R_{N}}e^{-\rho_{c}/2}{\rho_{c}}\ \int_{\Delta}^{\infty}dE\, f_{1}(E)(1-f_{2}(E+\hbar\omega))  \times\\ \nonumber &&  \qquad \left( \frac{E \left(E+\hbar\omega\right)-\Delta_{1}\Delta_{2}\cos \varphi +\left[ E \left(E+\hbar\omega\right) + \Delta_{1}\Delta_{2}\cos \varphi\right] S(t)/4}{\sqrt{\left(E^{2}-\Delta_{1}^{2}\right)\left(\left(E+\hbar\omega\right)^{2}-\Delta_{2}^{2}\right)}}\right) \\ \label{eq: rate complete formula-1}
\end{eqnarray}

In quantum information processors  the cutoff of the Fermi functions are well below the edge of the gap, so $f(E-\delta \mu_1)[1-f(E+\delta E -\delta \mu_{2})] \approx \exp[-(E-\delta \mu_1)/k_BT]$. In Josephson junction with two reservoirs we assume the two chemical potential shifts are the same. If the junction is between a large lead and a small island the tunnel rates are dominated by the quasiparticle density in the leads rather than in the island and here we assume the label 1 is assigned to the larger lead.  

Taking the integrals of all terms and simplifying the rate, after some calculus one can find the tunneling rate of nonequilibrium quasiparticle for arbitrary superconducting phase difference $\varphi$ across the junction and parity transition energy $\delta E $ in a junction made of the same  superconducting material:   
\begin{eqnarray}\nonumber
&& \Gamma_{1}[\varphi, T, \delta E ,S(t)]= \frac{1}{2e^{2}R_{N}}e^{\frac{\delta \mu}{kT}-\frac{\rho_{c}}{2}}{\rho_{c}}  \Delta e^{\xi-\frac{\Delta}{kT}}\times \\ \nonumber && \left\{ \frac{1-\cos\varphi}{2}K_{0}\left(\xi\right) +\xi\frac{kT}{\Delta}\left(\frac{3+\cos\varphi}{4}+\frac{9-\cos\varphi}{32}\frac{kT}{\Delta}\right)K_{1}\left(\xi\right)\right.\\ \nonumber && \left. +\left[\frac{1+\cos\varphi}{2}K_{0}\left(\xi\right)+\xi\frac{kT}{\Delta}\left(\frac{3-\cos\varphi}{4}+\frac{9+\cos\varphi}{32}\frac{kT}{\Delta}\right)K_{1}\left(\xi\right)\right]\frac{S(t)}{4}\right\}\\ 
\label{eq. appA final} 
\end{eqnarray}

One limitation is to consider that the transition energy between the two parity states is of the order of the qubit energy. Therefore, in the limit of $\xi=\delta E /2kT\gg 1$  given that $\delta E /\Delta\ll 1$, from eq. (\ref{eq. appA final}) one can derive:

\begin{eqnarray}
 \nonumber  
 && \Gamma_{1}[\varphi, T, \delta E ,S(t)] = \frac{1}{2 e^{2}R_{N}}e^{\frac{\delta \mu}{kT}-\frac{\rho_{c}}{2}}\rho_{c} \sqrt{\frac{\pi}{2}}e^{-\frac{\Delta}{kT}} \times \\ \nonumber && \left\{ \left(\frac{3+\cos\varphi}{4}\right)kT\sqrt{\xi}+\left(\frac{1-\cos\varphi}{2}+\frac{3}{8}\left(\frac{3+\cos\varphi}{4}\right)\frac{kT}{\Delta}\right)\frac{\Delta}{\sqrt{\xi}}\right.\\ \nonumber 
&+&\left.\left[\left(\frac{3-\cos\varphi}{4}\right)kT\sqrt{\xi}+\left(\frac{1+\cos\varphi}{2}+\frac{3}{8}\left(\frac{3-\cos\varphi}{4}\right)\frac{kT}{\Delta}\right)\frac{\Delta}{\sqrt{\xi}}\right]\frac{S(t)}{4}\right\} \\
\label{eq. rate append}
\end{eqnarray}

In the low temperature limit and large superconducting gap, using the chemical potential shift in the main text,  one can consider the fluctuation of phase is minimal and therefore it is of the order of electromagnetic vacuum fluctuations. By substitution we can  further simplify the tunnelling rate of eq. (\ref{eq. rate append}) into:
\begin{eqnarray}\nonumber
 &&\Gamma_{1}[\varphi, \delta E , T=0]=\frac{ \sqrt{\frac{\Delta E_c}{ \delta E  E_J }}}{8e^{2}R_{N}} \frac{n_{nqp}}{D_i(E_{f})}\  e^{-\sqrt{\frac{E_c}{8E_J}}} \times  \\ \nonumber  && \left\{ \left(1+\frac{3\delta E }{4\Delta} \right)\left(1+\frac{1}{4}\sqrt{\frac{E_c}{2E_J}}\right) -  \left(1 -\frac{\delta E }{4 \Delta} \right) \left(1-\frac{1}{4}\sqrt{\frac{E_c}{2E_J}}\right) \cos \varphi \right\}\\
\label{eq. appA final}
\end{eqnarray}

 \section{Crossover coupling $\rho^*(T)$}
 
 In order to find the condition for existence of the minimum tunneling rate at low temperature let us take a partial derivative of the tunneling rate with respect to the temperature. Simplifying the equation provides  the crossover environmental fluctuations

\begin{eqnarray}
\rho^*(T)=\frac{\left( 1+\frac{1}{2\xi}\right) K_1(\xi) -\frac{1}{2}\left(K_0(\xi)+K_2(\xi)\right) }{\left(\frac{2\Delta}{\delta E }-\frac{1}{4\xi}-\frac{1}{2}\right) K_1(\xi)+\left(\frac{1}{4}-\frac{\Delta}{\delta E \xi} -\frac{2 \Delta}{\delta E }\right)K_0(\xi)+\frac{1}{4}K_2(\xi)}
\label{eq. rho general} 
\end{eqnarray}

Note that the crossover $\rho$ depends nontrivially on the superconducting phase. If one uses eq. (\ref{eq. appA final}) the phase dependence of the minimum rate can be found.

%
%
%

\end{document}